\documentstyle[preprint,aps,eqsecnum]{revtex}                
\begin{document}
\draft
\title{The Magnetic Field Spectrum in a Plasma in Thermal 
Equilibrium in the Epoch of Primordial Nucleosynthesis}
\author{Merav Opher\footnote{email: merav@orion.iagusp.usp.br}, 
Reuven Opher\footnote{email: opher@orion.iagusp.usp.br}}
\address{Instituto Astron\^omico e Geof\' \i sico - IAG/USP, Av. Miguel 
St\'efano, 4200 \\
CEP 04301-904 S\~ao Paulo, S.P., Brazil}

\date{\today}     

\maketitle

\begin{abstract}
The low frequency magnetic field spectrum in the primordial plasma is of 
particular interest as a possible origin of magnetic fields in the universe 
(e.g., Tajima et al. 1992 and Cable and Tajima 1992).
We derive the magnetic field spectrum in the primordial plasma, 
in particular, at the epoch of primordial nucleosynthesis.
The pioneering study of Cable and Tajima (1992) of the electromagnetic 
fluctuations, based on the {\it Fluctuation-Dissipation Theorem}, is extended. 
Our model describes both the thermal and collisional effects in a plasma. 
It is based on a kinetic description with the BGK collision term. 
It is shown that the zero-frequency peak found by Cable and Tajima (1992) 
decreases. At high frequencies, the blackbody spectrum is obtained 
naturally without the necessity of the link procedure used by them. At
low frequencies ($\omega \leq 4\omega_{pe}$, where $\omega_{pe}$ 
is the electron plasma frequency) it is shown that the magnetic field 
spectrum has more energy than the blackbody spectrum in vacuum.  
\end{abstract}

\section{Introduction}
\label{sec:In}
Although plasma is the main constituent of the primordial universe, very
few previous studies deal directly with plasma phenomena.
The effect of a plasma in cosmology normally has been studied with respect to 
the origin of the magnetic field. For example, the study of Harrison 
\cite{har} elaborates a model of the origin of the magnetic field 
due to turbulence in the primordial plasma. There have been studies 
\cite{mag} that analyze the effect of a magnetic field on primordial 
nucleosynthesis. Other studies, like that of Halcomb et al. 
\cite{hal}, deduced the dispersion relation of waves, taking into account 
the expansion of the universe. There still lacks a general study of plasma 
phenomena related to cosmology.  

A plasma in thermal equilibrium, sustains fluctuations of the magnetic 
field (even for a non-magnetized plasma). This study is concerned with the 
study of the magnetic field spectrum in the primordial plasma. 

The electromagnetic fluctuations in a plasma has been made in numerous works, 
including those of Dawson \cite{daw}, Rostoker et al.\cite{ros}, 
Sitenko et al. \cite{sit1}, and Akhiezer \cite{akh1}. Most of the results are 
compiled in the books of Sitenko and Akhiezer et al. \cite{sit2,akh2}. 
Little attention has been given to the question of how the magnetic field 
spectrum looks in a plasma. A naive answer to this question might be that it 
is a blackbody spectrum with a cut-off at the plasma frequency, knowing that 
photons only propagate in a plasma for $\omega > \omega_{p}$, where 
$\omega_{p}$ is the plasma frequency. This is not true however, due to the     
magnetic fluctuations of the plasma.

Cable and Tajima and Tajima et al. \cite{tc1,tc2,tc3} performed a broad 
study of the magnetic field fluctuations in a plasma. They based 
their analyses on the {\it Fluctuation-Dissipation Theorem}. They were 
concerned, in particular, with the low-frequency spectrum of fluctuations, 
because, as they pointed out, no expression exists. 

The {\it Fluctuation-Dissipation Theorem} \cite{sit2,akh2} predicts the 
intensity of electromagnetic fluctuations. The intensity of such fluctuations 
is highly dependent on how the plasma is described, in particular, on the 
dissipation mechanisms used.

Cable and Tajima and Tajima et al. \cite{tc1,tc2,tc3} studied the 
magnetic field fluctuations, for several cases. Two of their descriptions 
concern the primordial plasma which we are interested in, which is an 
isotropic, non-magnetized and non-degenerate plasma: a) a cold, gaseous 
plasma and b) a warm, gaseous plasma described by kinetic theory. 

In their study, Cable and Tajima (hereafter CT) in case (a) used the {\it cold 
plasma} description with a constant collision frequency. In case (b) they 
analyzed the spectrum of fluctuations only for low frequencies, with 
the {\it warm plasma} description for phase velocity $\omega/k$ less or equal 
to the thermal velocity of the electrons, $v_{e}$ and the ions, $v_{i}$ in 
a collisionless description.

For the {\it cold plasma} description the spectrum that they obtain has 
a large zero-frequency peak. As the frequency is increased, the spectrum first 
drops below the blackbody spectrum in vacuum, then becomes the blackbody 
spectrum at high frequencies. In case (b), for the {\it warm plasma} 
description, the analyses was made only for the low frequency regime and 
they argued that the zero-frequency peak is present as well. They argue 
that the energy contained in the peak is approximately equal to the energy 
{\it lost} by the plasma cut-off effect. 
 
In order to obtain a correct magnetic field spectrum, it is 
necessary to describe the plasma in the most complete way as possible,
taking into account thermal and collisional effects in a unified
description.

In this study we extend the pioneering work of CT, presenting a model that 
includes, in a unified description, collisional and thermal effects. 
Our model is based on kinetic theory incorporating thermal effects for all 
frequencies and wave numbers (not only for 
$\omega/k \leq v_{e},~v_{i}$). In order to describe the collisions 
that exist in the plasma, we used a model collision term. This collision term 
describes binary collisions, as used in the work by CT. In this 
way, we extend the previous model describing thermal and collisonal effects 
for all frequencies and wave numbers. Their description, the {\it cold plasma} 
and the {\it warm plasma} description in the collisionless case, are special 
cases of this model. (A pure collisionless treatment is unreal, such as case 
(b) extended to all frequencies, since if there were no collisions, then 
only Cherenkov emission could produce fluctuations, and there are no 
particles traveling fast enough to emit light waves.) 

However, for a fully ionized plasma as is our case, a treatment that takes 
into account collisions in a more complete way is necessary. 
Our model, an extention of the CT model, describes the 
basic features of a kinetic description. 

We present in Section \ref{sec:mag} the general expressions of the 
magnetic field fluctuations based on the {\it Fluctuation-Dissipation Theorem}. 
We review the {\it cold plasma} description in Section \ref{sec:co} and the
{\it warm plasma} description in the collisionless case, in Section 
\ref{sec:war}. In Section \ref{sec:dis} we present a general discussion and 
criticism of the assumptions made by CT. 
In Section \ref{sec:mod} we present our model. Finally, in Section 
\ref{sec:con}, we discuss the results and present our conclusions.

\section{Magnetic Field Fluctuations}
\label{sec:mag}

The spectrum of fluctuations of the electric field in a plasma, given
by the {\it Fluctuation-Dissipation Theorem} (for the deduction of 
the {\it Fluctuation-Dissipation Theorem} from the general relation of 
fluctuations in a plasma see \cite{sit2,akh2}) is,  

\begin{equation}
{\frac{1}{8\pi} {\langle E_{i}E_{j} \rangle}_{{\bf k}{\omega}}=
\frac{i}{2}\frac{\hbar}{e^{{\hbar}{\omega}/T}-1}(\Lambda_{ji}^{-1}}-
{\Lambda_{ij}^{-1*}})~,
\label{ki}
\end{equation}
where
\begin{equation}
\Lambda_{ij}({\omega},{\bf k})=\frac{k^{2} c^{2}}{{\omega}^{2}}
\left( \frac{k_{i}k_{j}}{k^2}-{{\delta}_{ij}} \right) +
{\varepsilon}_{ij}\left( {\omega},{\bf k}\right) ~,
\label{lam}
\end{equation}
where $\varepsilon_{ij}({\omega,{\bf k}})$ is the dielectric tensor of the
plasma.
For an isotropic plasma,
\begin{equation}
\Lambda_{ij}=\frac{k_{i}k_{j}}{k^2}{\varepsilon}_{L}+
\left( {\delta}_{ij}-\frac{k_{i}k_{j}}{k^2}\right) \left( 
{\varepsilon}_{T}-\frac{k^2 c^2}{{\omega}^2}\right) ~,
\end{equation}
where $\varepsilon_{L}$ and $\varepsilon_{T}$ are, respectively, the 
longitudinal and transverse dielectric permittivities of the plasma. 
In this case \cite{sit2},
\begin{equation}
{\langle E_{i}E_{j} \rangle}_{{\bf k}{\omega}}=8\pi
\frac{\hbar}{e^{{\hbar}{\omega}/T}-1}{\left\{ \frac{k_{i}k_{j}}{k^2}
\frac{Im~{\varepsilon}_{L}}{{\mid {\varepsilon}_{L} \mid}^{2}}+
\left( {\delta_{ij}}-\frac{k_{i}k_{j}}{k^2} \right )
\frac{Im~{\varepsilon}_{T}}
{{\mid {\varepsilon}_{T}-{(\frac{k c}{\omega})}^{2} \mid}^2}\right\} }~.
\label{pa}
\end{equation}
Using the fact that ${\bf B}_{{\bf k}{\omega}}=(c/w){\bf k} \times
{\bf E}_{{\bf k}{\omega}}$, we have the expression for the magnetic
field fluctuations,

\begin{equation}
{\langle B_{i}B_{j} \rangle}_{{\bf k}{\omega}}=8\pi
\frac{\hbar}{e^{{\hbar}{\omega}/T}-1}\left( \delta_{ij}-
\frac{k_{i}k_{j}}{k^2} \right )
{\left( \frac{k c}{\omega}\right) }^{2}
\frac{Im~{\varepsilon}_{T}}
{{\mid {\varepsilon}_{T}-{\left(\frac{k c}{\omega}\right)}^{2} \mid}^2}~.
\label{kin}
\end{equation}

So, 
\begin{equation}
\frac{{\langle B^{2} \rangle}_{{\bf k}{\omega}}}{8\pi}=2
\frac{\hbar}{e^{{\hbar}{\omega}/T}-1}
{\left( \frac{k c}{\omega} \right) }^{2}
\frac{Im~{\varepsilon}_{T}}
{{\mid {\varepsilon}_{T}-{\left( \frac{k c}{\omega}\right) }^{2} \mid}^2}~.
\label{magn}
\end{equation}

An intuitive way to understand the above expression, is that the 
{\it Fluctuation-Dissipation Theorem} takes into account the emission 
and absorption processes in a plasma, and knowing that in 
equilibrium they are equal, the fluctuation 
level is evaluated. As Tajima et al. (1992) point out \cite{tc3}, 
``an individual mode decays by a certain dissipation, giving up energy to 
particles or other modes, while particles (or other modes) excite new modes 
and repeat the process and the amount of fluctuations is related to the 
dissipation''. 

In short, to determine the fluctuations of the magnetic field in a plasma in 
equilibrium or quasi-equilibrium, it is sufficient to know the 
transverse dielectric permittivity of the plasma, in particular, the 
dissipation mechanisms present for each frequency and wave number 
($Im~\varepsilon_{T}$). This depends on the treatment used to describe the 
plasma. 

Another important feature of the magnetic field spectral distribution, can 
be seen from Eq. (\ref{magn}). The equation 
\begin{equation}
\varepsilon_{T}(\omega,{\bf k})-{\left ( \frac{kc}{\omega} \right)}^{2}=0
\end{equation}
determines the transverse eigenfrequencies of the plasma. Therefore, the 
magnetic field spectral distribution has steep maximae 
at the frequencies that correspond to the transverse plasma 
eigenfrequencies.
In the transparency region ($Im~\varepsilon_{T} \ll Re~\varepsilon_{T}$), 
the magnetic spectrum has $\delta$-function-like maximae near the 
eigenfrequencies (i.e., the frequency spectrum of the fluctuations contains 
only the transverse eigenfrequencies in the plasma, \cite{akh2}).
Knowing that the photons have the dispersion relation in the plasma 
$\omega^{2}=\omega_{p}^{2}+k^{2}c^{2}$, we note that for frequencies 
$\omega \gg \omega_{p}$ the eigenfrequencies manifest themselves 
and the magnetic field spectrum behaves like a blackbody spectrum.

\subsection{Cold Plasma Description}
\label{sec:co}

The {\it cold plasma} description does not take into account the thermal 
movement of the electrons, describing them by fluid equations. In such a 
description, collisionless damping is not included. To include collisions 
in this case is straightforward, being only necessary to add a new term 
$\propto \eta v$ in the fluid equations, where $\eta$ is the collision 
frequency and $v$ the velocity of the particles. 
We follow here the CT study.

CT used a multifluid model of the plasma (neglecting the ${\bf v \times B}$ 
forces due to the smallness of the velocities and the 
electromagnetic fields),
\begin{equation}
m_{\alpha}\frac{d{\bf v}_{\alpha}}{dt}=e_{\alpha}{\bf E}-
\eta_{\alpha}m_{\alpha}{\bf v}_{\alpha}~,
\label{m}
\end{equation}
where $\alpha$ is a particle species label and $\eta_{\alpha}$ is the 
collision frequency of species $\alpha$. 
From the above equation (performing a Fourier transformation and 
re-arranging the terms), the dielectric tensor can be obtained: 
\begin{equation}
{\varepsilon}_{ij}({\omega},{\bf k})={\delta}_{ij}- 
\sum_{\alpha} \frac{\omega_{p\alpha}^{2}}{\omega(\omega+i\eta_{\alpha})}
\delta_{ij}~.  
\label{di}
\end{equation}
CT studied the case of an electron-positron plasma. This is the plasma that 
dominated the universe in the beginning of primordial nucleosynthesis, at 
$T\sim 1~MeV$. In the case of an electron-positron plasma, 
${\omega}^{2}_{pe^{+}}={\omega}_{pe^{-}}^{2}$ and 
$\eta_{e^{+}}=\eta_{e^{-}}=\eta$. Eq. (\ref{di}) then becomes
\begin{equation}
{\varepsilon}_{ij}({\omega},{\bf k})={\delta}_{ij}- 
\frac{\omega^{2}}{\omega(\omega+i\eta)}\delta_{ij}~,  
\end{equation}
where ${\omega_{p}}^{2}={\omega_{pe^{+}}}^{2}+{\omega_{pe^{-}}}^{2}$.       
$\eta_{e}$ was taken as the Coulomb collision frequency, 
${\eta}_{e}=2.91\times 10^{-6}~n_{e}~ln{\Lambda}T^{-3/2}(eV)~s^{-1}$, where 
$n_{e}$ is the electron (positron) density. 
(For the case of an electron-proton plasma, also treated by CT, 
${\eta}_{p}=4.78\times 10^{-18}~n_{e}~ln{\Lambda}T^{-3/2}(eV)~s^{-1}$). 
That is, this collision frequency describes the binary collisions in 
a plasma. 

For an isotropic plasma, the transverse dielectric permittivity is,
\begin{equation}
\varepsilon_{T}(\omega,{\bf k})=1-\frac{\omega_{p}^{2}}{{\omega}
(\omega+i\eta)}~.
\label{co2}
\end{equation}
Substituting Eq. (\ref{co2}) in Eq. (\ref{magn}), CT obtained the magnetic 
field spectrum in the {\it cold plasma} description. If relativistic 
temperatures effects are included, the substitution 
$\omega_{p} \rightarrow \omega_{p}/\sqrt{\gamma}$ is made.

The magnetic field spectrum ${\langle B^{2} \rangle}_{\omega}$ is found 
by integrating ${\langle B^{2} \rangle}_{k\omega}$ over wave numbers $k$ (and 
dividing by $(2\pi)^{3}$). The ${\langle B^{2} \rangle}_{\omega}$ diverges 
for high wave numbers. CT deal with this problem, breaking the integration on 
wave number into two intervals.  One interval runs from $\mid{\bf k}\mid =0$ to 
$\mid{\bf k}\mid= k_{cut}$. The other interval runs from 
$\mid{\bf k}\mid=k_{cut}$ to $\mid{\bf k}\mid=\infty$. In the first interval, 
they keep the collision frequency $\eta$ finite and in the
second interval they let $\eta \rightarrow 0$ and drop the 
low-frequency part of the spectrum. The final expression obtained was
\begin{eqnarray}
\frac{{\langle B^{2} \rangle}_{\omega}}{8{\pi}} & = &\frac{1}{\pi^{2}}
\frac{{\hbar}{\omega^{'}}}{e^{({\hbar}\omega_{pe}/T)\omega^{'}}-1}
2{\eta^{'}}{\left( \frac{\omega_{pe}}{c} \right )}^{3} 
\int_{0}^{x_{cut}}dx \frac{x^{4}}{({{\omega}^{'}}^{2}+
{{\eta}^{'}}^{2})x^{4}+...}  \nonumber \\    
 & &  + \frac{\hbar{({{\omega}^{'}}^{2}-{{\omega_{p}}^{'}}^{2})}^{3/2}}
{2\pi\left ( {e^{({\hbar}\omega_{pe}/T)\omega^{'}}-1}\right ) }
{\left( \frac{\omega_{pe}}{c} \right )}^{3} 
\Theta[\omega-\omega_{hev}]~,
\label{meg}
\end{eqnarray}
where $\Theta$ is the Heaviside step function, 
$\omega^{'}=\omega/\omega_{pe}$, ${\omega_{p}}^{'}=\omega_{p}/\omega_{pe}$, 
$\eta^{'}=\eta/\omega_{pe}$ and $x=kc/{\omega_{pe}}$.
The first term extends up to the frequency 
$\omega_{hev}=\sqrt{{k^{2}_{cut}c}^{2}+\omega_{p}^{2}}$. The second 
term is the high-frequency and high-wave number expression (i.e., the  
spectrum for frequencies $\omega \geq \omega_{hev}$). 

The justification given to this break-up procedure was that $\eta$ should 
vanish smoothly as $k \rightarrow \infty$ and as long as the results ``do not 
critically depend on the manner in which $\eta$ approaches zero'' the abrupt 
cut-off should be acceptable as a crude model. No strong justification was 
given to the choice of $k_{cut}$. They chose 
$x_{cut} \equiv k_{cut}c/{\omega_{pe}} \cong 1$ mainly because this is the 
value that makes the frequency spectrum be smooth at the joining of 
the low-frequency and the blackbody spectrum \cite{tc1,tc3}. 

\subsection{Warm Plasma Description}
\label{sec:war}
The {\it warm plasma} description describes the plasma based on the kinetic 
theory that takes into account the thermal distribution of the particles. 
In this description collisionless damping, like Landau damping, appears.
The kinetic theory is based on the BBGKY hierarchy equations that describe 
a system of many particles. These equations are solved by expanding the 
distribution function of the many particles in terms of the plasma parameter 
$g=1/n\lambda_{D}^{3}$, where $\lambda_{D}$ is the Debye length and $n$ is 
the particle density. The description that is used usually is the 
collisionless description, based on the Vlasov equation that does not take 
into account collisions (and neglecting ${\bf v \times B}$ forces): 
\begin{equation}
\left ( \frac{\partial}{\partial t}+{\bf v}\cdot \nabla
+\frac{q}{m}{\bf E}\cdot {\nabla}_{\bf v}  \right ) f({\bf x},{\bf v},t)=0~,
\label{vla}
\end{equation}
where $\nabla \equiv \partial/\partial{\bf x}$ and 
$\nabla_{\bf v} \equiv \partial/\partial{\bf v}$.
The Vlasov equation in first order (in $g$) takes into account collisions, 
where the term on the right hand side of the equation is now 
the {\it collision term}:
\begin{equation}
\left ( \frac{\partial}{\partial t}+{\bf v}\cdot \nabla
+\frac{q}{m}{\bf E}\cdot {\nabla}_{\bf v}  \right ) f({\bf x},{\bf v},t)=
\left ( \frac{\partial f}{\partial t} \right ) _{C}~.
\label{vla1}
\end{equation}

For the collisionless description (assuming an isotropic plasma), the 
transverse dielectric permittivity is obtained from Eq. (\ref{vla}). Assuming 
that the particles have a Maxwellian velocity distribution, the transverse 
dielectric permittivity is \cite{kra,mel,ich,cle},
\begin{equation}
\varepsilon_{T}({\omega},{\bf k})=1- {\sum_{\alpha}} 
\frac{{{\omega}_{p\alpha}}^2}{{\omega}^2}
\left \{ \phi \left ( \frac{\omega}{\sqrt{2}k{v_{\alpha}}} \right ) -
i{\left ( \frac{\pi}{2} \right ) } ^{1/2}\frac{\omega}{kv_{\alpha}}
{exp \left ( -\frac{{\omega}^2}{2k^2{v_{\alpha}^2}}\right ) } 
\right \} ~,
\label{et}
\end{equation}
where $\alpha$ is the label for each species of the plasma and $v_{\alpha}$ 
is the thermal velocity for each species, $v_{\alpha}=\sqrt{T/m_{\alpha}}$.
$z={\omega}/\sqrt{2}kv_{\alpha}$ and 
$\phi (z)=2ze^{-{z^2}}\int_{0}^{z}e^{x^2}dx$. 

In order to investigate thermal effects, CT used the collisionless 
description (i.e., the Vlasov equation in the regime where $\omega/k$ is less 
than the thermal speed of the plasma constituents). They studied a 
hydrogen plasma, thus the region investigated was 
${\omega}/{k} \leq v_{e},v_{i}$ where $v_{e}=\sqrt{T/m}$ and 
$v_{i}=\sqrt{T/M}$. (The extension to an electron-positron plasma, as in 
case (a), is straightforward: $M \rightarrow m$ so 
$v_{i} \rightarrow v_{e^{+}}$). They used, therefore, 
Eq. (\ref{et}) in the limit $z \ll 1$ and expanded the plasma dispersion 
function $\phi(z)$. Substituting the approximate expression of 
$\varepsilon_{T}(\omega,{\bf k})$ (expanded in the limit of $z \ll 1$) 
in Eq. (\ref{magn}), they obtained ${\langle B^{2} \rangle}_{k\omega}/8\pi$. 

CT examined only the low frequency behavior of 
${\langle B^{2} \rangle}_{\omega}$, noting that it diverges for high wave 
numbers and a cut-off procedure is necessary. They assumed the same upper 
limit as in case (a), $x_{upper}=x_{cut} \cong 1$.

\subsection{Discussion and Criticisms}
\label{sec:dis}

For the {\it cold plasma} description, used in case (a), CT obtained a 
large zero-frequency peak in the magnetic field fluctuation spectrum 
(arguing that the {\it warm plasma} description exhibits it as well).

The total spectrum is obtained with the link of the low frequency term to 
the high frequency term (Eq. (\ref{meg})). The behavior of the spectrum can 
be seen in Figure 1a. We study an electron-positron plasma at 
$T=7\times10^{9}~K$ and $n_{e}=4.6\times 10^{30}~cm^{-3}$. This is the plasma 
in the beginning of primordial nucleosynthesis. The magnetic field spectrum 
$S({\omega})\equiv {\langle B^{2} \rangle}_{\omega}/{8\pi}$ is divided by a 
normalization $S_{0}=\omega_{pe}^{2}k_{B}T/c^{3}$. The dashed curve 
is the first term of Eq. (\ref{meg}), the low frequency spectrum. The 
dash-dot-dash curve is the second term of Eq. (\ref{meg}), the high frequency 
spectrum (obtained with $\eta \rightarrow 0$). We note the link point used 
between the two curves. The blackbody spectrum in vacuum is also plotted 
(solid curve). It can be seen that the general behavior is that after the peak, 
with increasing frequency, the spectrum drops below the blackbody spectrum. 
At high frequencies, it merges into the blackbody spectrum. In Figure 1b we 
plotted only the {\it cold plasma} 
spectrum for low and high frequencies. In Figure 1c is a electron-proton plasma 
at $T=10^{9} K$ and $n_{e}=5.4\times 10^{26}$. In the epoch 
of primordial nucleosynthesis, at lower temperatures, the 
electrons and positrons annihilate and the plasma is reduced to a plasma of 
protons and electrons. In Figure 1d we used the same plasma as in Figure 1c, 
but plotted only the {\it cold plasma} spectrum for low and high 
frequencies.

CT argued that the results do not critically depend on the upper limit. 
We show below that this is not true. As CT noted, the 
divergence occurs due to the subtle interaction between matter and 
radiation in small scales. CT used a classical fluid equation with a 
constant collision frequency (Coulomb collision frequency). This collision 
frequency describes the binary interactions in the plasma. They chose the 
cut-off $x_{cut} \cong 1$, basically because this is the value that makes the 
frequency spectrum smooth at the joining of the low-frequency and the 
blackbody spectrum.

In section VII of CT, they gave a quantum-mechanical
justification of not extending $k$ beyond $k_{cut}$. They argue that for 
$(\hbar k)^2/2m \gg k_{B}T$ the plasma has a negligible 
effect on the electromagnetic spectrum. Let us call this k, $k_{lim}$ 
(${(\hbar k_{lim})}^{2}/2m=k_{B}T$).  

When treating Coulomb collisions, a cut-off has to be taken, since at 
small distances the energy of the Coulomb interactions of the particles exceeds 
their kinetic energy which violates the applicability of the condition of the 
perturbation expansion (in the plasma parameter $g \ll 1$). 
This occurs approximately for distances $r_{min} \sim  e^{2}/T$ or, 
more exactly, for the distance of closest approach between a test particle and an electron 
in a plasma, $k_{max}=1/r_{min}\cong {Mmv^{2}}/{(m+M)} \mid eq\mid $, where 
$M$, $v$ and $q$ are respectively, the mass, velocity and charge of the test 
particle [15].

Comparing the value of $k_{cut}$, $k_{lim}$ and $k_{max}$
we see that $k_{cut}$ ($x_{cut}=k_{cut}c/\omega_{pe}\cong 1$)
is much smaller than the others. For example, for the
cases that CT used: 1) $T=10^{10}~K$, 
$n_{e}=4.8\times10^{30}~cm^{-3}$; 2) $T=10^{6}~K$, 
$n_{e}=6.5\times10^{9}~cm^{-3}$; and 3) $T=10^{4}~K$, 
$n_{e}=6.5\times10^{3}~cm^{-3}$ we found:   
1) $x_{lim}\equiv k_{lim}c/\omega_{pe} = 11.5$ and $x_{max}
\equiv k_{max}c/\omega_{pe} = 2444.4$;  
2) $x_{lim}=3.1\times10^{9}$ and $x_{max}=6.6\times10^{9}$; and 
3) $x_{lim}=3.1\times10^{10}$ and $x_{max}=6.6\times10^{11}$. 

As $x_{cut}$ is quite arbitrary, we decided to vary the upper limit in the 
first term in Eq. (\ref{meg}), until $x_{upper}=x_{max}$ to observe the 
behaviour of the curves. This is shown in Figure 2 where again we plot 
$S({\omega})/S_{0}$ vs $\omega/\omega_{pe}$. 

We plotted the low frequency spectrum term not only until 
$\omega_{hev}=\sqrt{{k^{2}}_{cut}c^{2}+{\omega_{p}}^{2}}$, as CT did, but 
extended it to higher frequencies.
The dash-dot-dash curve has the upper limit $x_{upper}=x_{cut}$ as CT used. 
The dotted curve has the upper limit $x_{upper}=2~x_{cut}$; the dashed curve 
has the upper limit $x_{upper}=5~x_{cut}$; and the dash-two dot-dash curve 
has the upper limit $x_{upper}=x_{max}$. We also plotted the blackbody 
spectrum in vacuum (solid curve).
It can be seen that by extending the upper limit to higher and higher values 
we obtain more and more of the blackbody spectrum, before the spectrum drops. 
This is easy to understand since, as we noted before, the magnetic field 
spectrum, obtained from the {\it Fluctuation-Dissipation Theorem}, contains 
the transverse eigenfrequencies of the plasma, the photons. When we extend 
the value of the upper limit, we permit higher eigenfrequencies to manifest 
themselves. This occurs up to the frequency when, due to the upper limit 
chosen, no more photons can manifest themselves.

Another feature that appears is that by increasing the upper limit the valley 
that appears, due to the plasma cut-off effect, becomes smaller and 
smaller. Eventually, for $x_{upper}=x_{max}$, we obtain a large peak for 
$\omega \sim 0$ but without any valley. In fact, for frequencies 
$\omega \leq 2 \omega_{pe}$ the curve is above the blackbody spectrum in 
vacuum. This can be seen in Figure 3a, where the dashed curve is the low 
frequency term (in Eq. (\ref{meg})) extended to high frequencies (not only 
for $\omega \leq \omega_{hev}$) with $x_{upper}=x_{max}$ compared to the 
blackbody spectrum in vacuum (solid curve). In Figure 3b we plotted only 
the {\it cold plasma} curve, extended to high frequencies, showing that 
with $x_{upper}=x_{max}$ the full blackbody spectrum is reproduced. 

CT argued, based on the behavior of the 
{\it cold plasma} spectrum with $x_{upper}=x_{cut}$, that the energy under 
the $\omega \sim 0$ peak is approximately equal to the energy stolen from 
the blackbody spectrum due to the plasma cut-off. In their words, this 
happens because, the ``plasma squeezes the fluctuation energy of modes with 
frequency less than $\omega_{p}$ into modes with frequencies very close to 
zero''.

First of all, clearly, this does not happen if the upper limit is high 
enough in order to reproduce the full blackbody spectrum at high 
frequencies (with $x_{upper}=x_{max}$) because the entire curve is above 
the blackbody spectrum. Second, we noted previously that a blackbody (photon) 
spectrum with a cut-off $\omega< \omega_{p}$ is expected taking into account 
that the photons only propagate for $\omega > \omega_{p}$ in a plasma, where 
$\omega_{p}$ is the plasma frequency. However, when we speak about blackbody, 
we speak about modes which propagate, in our case, photons. In vacuum, the 
photons have a dispersion relation $\omega=kc$ and they are present in the 
entire frequency spectrum. In a plasma, when the dispersion relation is 
$\omega^{2}={\omega_{p}}^{2}+k^{2}c^{2}$, they only appear for 
$\omega > \omega_{p}$.

Dawson \cite{daw} deduced the radiation spectrum in a
plasma by a Gedanken experiment. A slab of plasma at a temperature T is put 
between two blackbodies at temperature T. Between the plasma and the 
blackbodies are vacuum regions. Radiation is emitted by the blackbodies and 
enters the plasma. In equilibrium, the plasma radiates the same amount of 
radiation that it absorbs. Dawson deduced the density of radiation in a 
plasma as:
\begin{equation}
U_{p}=\left ( 1-\frac{{\omega_{p}}^{2}}{\omega^{2}} \right ) U_{v}~,
\end{equation}
where $U_{p}$ is the radiation density in the plasma and $U_{v}$ the blackbody 
radiation density in the vacuum. It can be seen that a new factor appears, 
due to the presence of the plasma. This kind of Gedanken experiment tells us, 
however, only about modes that propagate. Nothing is told about the modes 
that do not propagate which appear due to correlations in the plasma. 
They are present at low frequencies, $\omega < \omega_{p}$, but 
they also are present at $\omega > \omega_{p}$, contributing, besides 
the photons, to the magnetic field spectrum. Our results show that this 
happens only for very high frequencies ($\omega \gg \omega_{p}$) when 
the photons dominate the magnetic field spectrum. The magnetic field 
fluctuation spectrum has little to do with the photon blackbody spectrum in      
vacuum for $\omega < \omega_{p}$. In principle, the magnetic field spectrum 
can be greater, or less, than the photon blackbody spectrum in vacuum for 
$\omega < \omega_{p}$. The only manner to obtain the magnetic field spectrum 
is analyzing the magnetic field fluctuations from, for example, the 
{\it Fluctuation-Dissipation Theorem}. Using the {\it cold plasma} description, 
for example, with the cut-off $x_{upper} \sim x_{max}$, we find that the magnetic 
spectrum has more energy than the blackbody photon spectrum in vacuum.

\section{Our Model}
\label{sec:mod}

In order to extend the work of CT and have a more complete description 
of the plasma, we desire a model that includes thermal effects as well as 
collisional effects. For this, we need a kinetic description that 
takes into account collisions. We used the Vlasov equation in first order.
This equation gives collisional corrections to the Vlasov
equation in zero order (collisionless case). It retains terms of 
order $g$ and neglects terms of higher order. As the plasma parameter $g$ is 
much less than unity (for example, as pointed out by \cite{tc3}, 
in the epoch of $t=10^{-2}-10^{0}~s$ in the primordial universe, 
$g \cong 10^{-3}$), this is a good approximation, as terms of higher order  
are much smaller. As noted before, the term on the right hand side of 
Equation (2.14), $(\partial f/\partial t)_{C}$, is the {\it collision term}. 
Obtaining $(\partial f/\partial t)_{C}$ is a matter of great 
difficulty and different forms are required for various types of collisions, 
such as electron-electron, electron-neutral molecule, etc. Most of the 
expressions for $(\partial f/\partial t)_{C}$ involve integral functionals of 
the distribution function $f$. The basic difficulty in taking into account
the effect of collisions lies in the complexity of the solution of the kinetic
equation when the correct collision integral is used. The problem can be 
simplified if a {\it model} collision term is used, that is, an approximate 
expression.

The Boltzmann collision term takes into account all the possible binary collisions 
which the particle under observation might suffer. It is applied to weakly 
ionized plasma, when the scattering of charged particles by neutrals is 
predominant. In a fully ionized plasma (as in our case), the collisions are 
not predominantly binary and $(\partial f/\partial t)_{C}$ for a {\it test} 
particle does not derive mainly from the possibility that other particles 
approach very closely and abruptly deflect it. The cumulative effect of more 
distant particles is more important. The charged particles simultaneously 
interact with all the other particles in the Debye sphere, which is a large 
number. The Fokker-Planck collision term is appropriate for a fully ionized 
plasma. It is based in the notion that a large-angle deflection of a particle 
by collisions is produced more rapidly by a succession of small-angle 
scattering with distant particles. It takes into account the effect of 
microscopic fields produced by all the other particles in the 
plasma. A {\it test} particle then is subject to simultaneously ``grazing'' 
collisions and its progress in velocity space becomes a random walk. 

We used the BGK collision term as a rough guide to the inclusion of collisions 
in the plasma. BGK is a {\it model} equation of the Boltzmann collision term. 
(For a derivation see Clemnow et al. \cite{cle} and Alexandrov et al.
\cite{ale}.) As CT, we are then only treating binary
collisions. 
(Concerning this collision term, see for example the work of Sitenko and Gurin \cite{sit1}, 
who studied the effect of an effective binary collision frequency on the 
fluctuations in a plasma). A more complete treatment using the Fokker-Planck 
collision term is necessary. However, due to its complexity, as it involves integral 
functionals of the distribution function $f$, the kinetic equations are 
extremely difficult to solve. We used as an effective collision frequency, the 
Coulomb collision frequency as CT.

The BGK collision term conserves the number of particles. The universe at 
high temperatures, at the beginning of the nucleosynthesis ($T \sim 1~MeV$) 
is an electron-positron plasma. As it cools down ($T \leq 0.5~MeV$) the 
electron and positrons start to annihilate and finally, at low temperatures, 
the plasma is reduced to a plasma of protons and electrons. Let us see, if 
this is a good approximation in the early universe. 
The rate of annihilation is given by $\Gamma \sim nv\sigma$, 
where $n$ is the density, $v$ the average velocity and $\sigma$ the 
average cross section.  The average cross section is given by, 
$\sigma=\frac{2\pi{\alpha}^{2}}{s\beta}[\frac{(3-{\beta^4})A}{2\beta}-2
+{\beta^2}]$, where $\beta^{2}=1-\frac{4m^{2}}{s}$,
$A=ln[\frac{1+\beta}{1-\beta}]$, $\alpha=\frac{1}{137}$ and
$s={(2E)}^{2}$ with $m$ the electron mass and $E$ the electron energy
\cite{ll}. Comparing $\Gamma$ with the collision frequency $\eta$ (the Coulomb
collision frequency as used by CT), we see that
$\Gamma < \eta$. For example, for $T \sim 0.8~MeV$, 
$\Gamma=4\times 10^{16}~s^{-1}$ and $\eta=4\times 10^{17}~s^{-1}$.
Therefore, the characteristic time of the kinetic processes is shorter than for 
annihilation and the approximation of the conservation of
the number of particles at high temperatures is valid.
(As in our case, when the collision frequency is much less than the plasma
frequency, the collision time is the dominant time for the kinetic
processes.)
At low temperatures, an electron-proton plasma has to be considered. 

We note that because the electron-positron plasma strictly does not
conserve the particle number due to annihilation and creation, there may
not be a great advantage of using the BGK collision term, although the
above paragraph indicates that the annhilation frequency is very much smaller
than the collision frequency.

The plasma, in the epoch of primordial nucleosynthesis, is almost 
collisionless. (For example, for $T=10^{10}~K$, 
$\eta/\omega_{pe} \cong 10^{-3}$.) Therefore, a possible procedure that one 
might think of is to expand the dielectric permittivity in terms of 
$\eta/\omega$. However, we are interested in the spectrum for all 
frequencies, even $\omega \sim 0$, and no matter how small $\eta$ is, 
we require frequencies with $\omega < \eta$. Thus an expansion in 
$\eta/\omega$ cannot be made.

Therefore, we perform an analyses of the Vlasov equation in first order 
with the BGK collision term, without making any approximation,  
\begin{equation}
\left ( \frac{\partial f}{\partial t} \right )_{C} = -\eta(f-f_{max})~,
\end{equation}
where $\eta$ is the Coulomb collision frequency (considered constant). $f_{max}$ is  
given by $f_{max}({\bf x},{\bf v},t)=N({\bf x},t)f_{0}({\bf v})/{N_{0}}$ 
where the number density is $N=N_{0}+N_{1}$ and $f=f_{0}+f_{1}$, $f_{0}$ 
being the unperturbed Maxwellian distribution. Substituting this collision 
term in the equation of Vlasov in first order, Eq. (\ref{vla1}), performing a 
Fourier transformation and re-arranging the terms, we arrive at
\begin{equation}
f_{1}({\bf v})=\frac{1}{i({\omega}-i{\eta}-{\bf k\cdot v})}\left [ 
-\frac{e}{m}E_{i} \frac{\partial f_{0}}{\partial v_{i}} +
\eta \frac{N_{1}}{N_{0}}f_{0} \right ]~.
\end{equation}
To eliminate $N_{1}$ we integrate over velocity space \cite{cle},
\begin{equation}
N_{1}=\int f_{1} ({\bf v}) d{\bf v}~.
\end{equation}
We have the current ${\bf j}=e\int v f_{1}({\bf v}) d{\bf v}$. Knowing that 
$j_{i}=\sigma_{ij}E_{j}$, then for an isotropic plasma the transverse 
permittivity is easily obtained (generalizing for several species):
\begin{equation}
\varepsilon_{T}(\omega,{\bf k}) = 1 + \sum_{\alpha} 
\frac{{\omega_{p\alpha}}^{2}}{\omega^{2}}
\left ( \frac{\omega}{\sqrt{2}kv_{\alpha}} \right ) Z \left ( 
\frac{\omega+i\eta_{\alpha}}{\sqrt{2}kv_{\alpha}} \right )~,
\label{etw}
\end{equation}
where $\alpha$ is the label for each specie of the plasma, $v_{\alpha}$ 
is the thermal velocity for each specie and $Z(z)$ is the Fried $\&$ Conte 
function \cite{fc}, 
\begin{equation}
Z(z)=\frac{1}{\sqrt{\pi}}\int_{-\infty}^{\infty} \frac{dt~e^{t^2}}{t-z}~.
\end{equation}
If relativistic temperatures effects are included, the substitution 
$\omega_{p} \rightarrow \omega_{p}/\sqrt{\gamma}$ is made.

It can be seen that no approximation needs to be made on 
${\omega}/{\sqrt{2}kv_{\alpha}}$ and ${\eta}/{\sqrt{2}kv_{\alpha}}$. 
(Re-writing $Z(z)$ in terms of the error function, 
$Z(z)=i\sqrt{\pi}exp(-{\xi^{2}})[1+erf(i{\xi})]$ \cite{fc}, 
it can be easily solved numerically without having to take any asymptotic 
limit).

Another thing that we need to be careful of is related to the dielectric 
permittivities in the region of $\omega/k \geq c$, where $c$ is the velocity of 
light. For $\omega/k >c$, the term connected with Cherenkov emission in
the imaginary part of the dielectric permittivities has to vanish.
This is because there are no particles with velocities greater than the speed of 
light to produce Cherenkov emission.  
(The condition for Cherenkov emission of a wave $\omega(k)$ requires 
particles with velocities satisfying the condition $v=\omega/k \equiv v_{p}$, 
where $v_{p}$ is the phase velocity of the wave.)
For our case, a plasma in thermal equilibrium, we assume as CT in
section V of their article \cite{tc2}, that the distribution function is a 
Maxwellian one. The calculation of the permittivity tensor (and the 
dielectric permittivities) involves integral functionals of the distribution 
function.
As pointed out by Melrose \cite{mel}, the non-vanishing of the imaginary 
parts, in a collisionless treatment (that is Cherenkov emission) is due 
entirely to the effects of unphysical particles with $v>c$ in the Maxwellian 
distribution. We deal with this problem by requiring that the Cherenkov emission 
in the imaginary part is zero in the regime $\omega/k >c$. The term of the
imaginary part, connected with the collisional damping, is not set to zero.

As photons have phase velocities greater than the speed of light ($\omega/k >c$), the
Cherenkov emission cannot produce the photons. In a {\it pure collisionless} 
treatment, the imaginary part has to be set to zero in the regime of $\omega/k>c$.
In this case, no photons are produced and the treatment is unrealistic. 
In our model, which includes collisions and thermal effects, only the 
term connected with the Cherenkov emission has to be set to zero. In 
this treatment we ensure that the photons are produced only by collisions,
that is, by bremsstrahlung (free-free) emission.

Therefore, by forcing the Cherenkov
emission in the imaginary part to be zero in the regime of $\omega/k >c$, we 
ensure that no spurious effects contaminate the result. We do this 
not only in the regime of photons $(\omega > \omega_{pe})$, but 
for all frequencies. That is, Cherenkov emission of magnetic fluctuations 
occur only in physical regimes, i.e., for $\omega/k \leq c$. 
It is left for a future study a fully relativistic treatment (needed for 
the primordial universe at high temperatures). There, the imaginary parts 
of the dielectric permittivities are zero for $v_{\phi}>c$, where 
$v_{\phi}=\omega/k$, when $\gamma={(1-v_{\phi}^2/c^2)}^{-1/2}$ and 
$p_{\phi}=\gamma_{\phi}mv_{\phi}$ are imaginary.

In principle, the fact that the term connected with the Cherenkov emission, 
for $\omega/k >c$ is forced to be zero, could be troublesome. This procedure,  
however, did not change appreciably the results. For example, the change of
the intensity of the magnetic field fluctuations 
$\Delta {\langle B^{2} \rangle}_{\omega}$ (after integrating on wave number) 
for $T=7\times 10^{9}~K$, $n_{e}=4.6\times 10^{30}~cm^{-3}$ is:  
$10^{-7}$, $10^{-3}$ and $10^{-5}$ for $\omega/\omega_{p}=0.1,~1.0$ and 
$100.0$, respectively. The change is the difference between
${(Im \varepsilon_{T})}_{C} \neq 0$ and ${(Im \varepsilon_{T})}_{C} = 0$,
where ${(Im \varepsilon_{T})}_{C}$ is the term in the 
imaginary part of the transverse dielectric permittivity connected with
Cherenkov emission.

Before we go on, it is interesting to comment and emphasize that both 
the {\it cold plasma} description and the {\it warm plasma} 
collisionless description are particular solutions of this model. Let us 
consider, as CT did, an electron-positron plasma. 
For ${\mid z \mid}^{2} \gg 1$, where $z=({\omega+i\eta})/{\sqrt{2}kv_{e}}$, 
we expect to obtain the {\it cold plasma} approximation. In the limit of 
${\mid z \mid}^{2} \gg 1$, 
$Z(z)=-\frac{1}{z}-\frac{1}{2z^{3}}+...-i\sqrt{\pi}ze^{-z^{2}}$. 
The last term is due to Cherenkov emission, but as we noted before, in this 
limit this is a spurious solution and we need to set it to zero. Taking only 
the first term, and substituting in Eq. (\ref{etw}), we obtain the 
{\it cold plasma} dielectric permittivity, as expected. 

The {\it warm plasma} collisionless description is obtained by setting 
$\eta \rightarrow 0$. In this limit, if we write $z=x+iy$, 
$iy=i0$, $\bar{\phi}(x+i0)=\phi(x)-i\sqrt{\pi}xe^{-x^{2}}$ 
($Z(z)=-1/z~\bar{\phi}$), Eq. (\ref{etw}) is equal to the 
{\it warm plasma} collisionless dielectric permittivity, as expected.

We substitute the dielectric permittivity Eq. (\ref{etw}) in Eq. (\ref{magn}) 
obtaining the magnetic field spectrum ${\langle B^{2} \rangle}_{k\omega}$. 
Here, a divergence at high wave numbers also occurs. Let us discuss this in 
detail. Our model uses a kinetic 
theory description with a model collision term that describes the binary 
collisions in the plasma. In our case, a 
cut-off has to be taken, since for very small distances the energy of the Coulomb 
interactions of the particles exceeds their kinetic energy which violates the 
applicability of the condition of the perturbation expansion (in the plasma 
parameter $g \ll 1$). This occurs approximately for distances 
$r_{min} \sim  e^{2}/T$, or more exactly, the distance of closest approach 
between a test particle and an electron 
in a plasma, $k_{max}=1/r_{min}\cong {Mmv^{2}}/{(m+M)} \mid eq\mid $, where 
$M$, $v$ and $q$ are respectively, the mass, velocity and charge of the test 
particle \cite{ich}. 

The cut-off procedure can only be removed, treating properly 
the effects of distant encounters. This can be done with the Fokker-Planck collision term. 
Thompson and Hubbard, and Hubbard in several works \cite{th,h1,h2}, analyzed 
the Fokker-Planck equation and its coefficients. The diffusion and friction 
coefficients that appear in the Fokker-Planck equation, take into account 
correlation effects between distant particles in the plasma. When higher 
order terms in the Fokker-Planck equation are calculated and summed, a 
term resembling the Boltzmann collision term is obtained. In their treatment, 
they showed that the cut-off procedure is unnecessary.
However, due to the complexity of the solution of the kinetic equation with 
the Fokker-Planck collision term, we used the BGK collision term.  With this 
model collision term, a cut-off is necessary and we chose $x_{max}$ 
consistent with this model collision term. A more exact treatment, however, 
is needed.

In Figure 4a we plot the magnetic field spectrum 
$S({\omega})={{\langle B^{2} \rangle}_{\omega}}/{8\pi}$ (divided by the 
normalization $S_{0}=\omega_{pe}^{2}k_{B}T/c^{3}$) vs $\omega/\omega_{pe}$ 
for an electron-positron plasma at $T=7\times 10^{9}~K$ and 
$n_{e}=4.6\times10^{30}~cm^{-3}$. The dotted curve is our model and we 
compare it with the blackbody spectrum in vacuum (the solid curve). 
In Figure 4b, we  extend the curves to high frequencies, showing the 
behaviour of the blackbody at high frequencies. Figures 4c and 4d 
are for an electron-proton plasma, with $T=10^{9}~K$ and 
$n_{e}=5.4\times 10^{26}$.

As we commented before (section \ref{sec:dis}), the results are very dependent 
on the cut-off chosen. Using $x_{max}$ as the cut-off, we obtain results that 
differ from the CT results: 
First, the peak intensity found by them for frequencies $\omega \sim 0$ 
decreases. This is due to the kinetic plasma effects that smear out the peak.
However, it is interesting to see that qualitative agreement 
between the work of CT and ours exist with respect to 
the zero frequency peak; Second, we 
obtain the blackbody naturally for high frequencies; and Third, the magnetic 
field spectrum has more energy than the blackbody spectrum for frequencies 
$\omega \leq 4~\omega_{pe}$. 

\section{Conclusions and Discussion}
\label{sec:con}

The magnetic field spectrum can be deduced from 
the fluctuations of the magnetic field described by the {\it Fluctuation-
Dissipation Theorem} and it is highly dependent on the way the plasma is 
described. We discussed the {\it cold plasma} description and the 
{\it warm plasma} description in the collisionless case studied by CT. 
We showed that $x_{cut}$ is much smaller than $x_{max}$ and $x_{lim}$, 
where $x_{lim}=k_{lim}c/\omega_{pe}$ used in Sec. VII 
of CT (arguing that for $k > k_{lim}$ the plasma has a negligble effect). 
$x_{max}$ is the cut-off used in treating binary collisions, $(x_{max})^{-1}$ 
being the distance of closest approach between a test particle and an 
electron in a plasma (divided by $c/\omega_{pe}$).

We also showed that the {\it Fluctuation-Dissipation Theorem} contains the 
eigenfrequencies of the plasma (in the transverse case), the photons. 
Using the {\it cold plasma} description with the upper limit 
$x_{upper}=x_{max}$, we obtain the blackbody spectrum at high frequencies  
naturally, without the necessity of a {\it link} procedure used by CT. For this case, 
the valley disappears and the curve is above the blackbody spectrum 
in vacuum. 

The calculations were made for two types of primordial plasmas: The 
electron-positron plasma at the beginning of the Big Bang nucleosynthesis; 
and the electron-proton plasma at lower temperatures.

The manner to obtain the entire magnetic field spectrum is analyzing the 
magnetic field fluctuations. This is the only way to obtain information, 
not only about modes that propagate (i.e., photons), but also modes 
that do not propagate. The modes that do not propagate appear not only at 
low frequencies but also at high frequencies due to the correlations in the 
plasma. Only at very high frequencies does the photon contribution dominate 
the magnetic field spectrum. The argument used by CT, that the energy under 
the peak is almost equal to the energy ``stolen'' by the plasma cut-off effect 
of the blackbody in vacuum, is incorrect. There is no reason why we have to have 
the same energy as the blackbody spectrum in vacuum for photons for 
$\omega < \omega_{pe}$, since the photons have a different dispersion 
relation than the fluctuations in the plasma. In fact, using 
a upper limit $x_{upper}=x_{max}$ in the {\it cold 
plasma} description, for example, the spectrum obtained is above 
the blackbody spectrum in vacuum. 

The reason why the collective modes of the plasma can have more energy for 
$\omega \leq \omega_{p}$ than the photons in vacuum, can be 
understood as follows. Photons are massless bosons with the dispersion relation 
$\omega^{2}=k^{2}c^{2}$. For the energy interval, 
$0 \leq \omega \leq \omega_{p}$, the wave number interval is $k=0$ to $k$ 
equal to $\omega_{p}/c$. A relatively small amount of phase space is involved. 
For the collective motions of the plasma, in general, we have a larger amount 
of phase space. For example, for plasmons with energy 
$\omega \sim \omega_{p}$, the amount of phase space extends to a maximum $k$ 
of $k_{D} \cong \omega_{p}/v_{t}$, which is greater than $\omega_{p}/c$ for the 
photons. In general, for a given frequency 
for $\omega < \omega_{p}$, the greater phase available to the collective 
modes of the plasma (than that of the photons) implies more energy, or a higher 
spectrum. 

We presented a model that incorporates, in the same description, thermal 
and collisional effects. We used the Vlasov equation with 
the BGK collision term. This collision  
term describes the binary collisions in the plasma. A model that takes into 
account collisions in a more complete way is necessary. 
For a fully ionized plasma it is necessary to use the Fokker-Planck collision 
term that takes into account the effect of the microscopic fields.
Due to the complexity of the solution
of the kinetic equation with such a collision term, we used the BGK collision 
term. This model, an extention of the CT model, 
describes the basic features of a kinetic description.

As we noted before, the results are very dependent on the cut-off chosen. 
Using $x_{max}$ as the cut-off, consistent with the collision term used, we 
obtain results that differ from the CT results. The final 
magnetic spectrum of a non-magnetized plasma in thermal equilibrium has the 
following characteristics:
a) The peak intensity found by CT for frequencies $\omega \sim 0$ 
decreases; b) The blackbody is obtained naturally for high 
frequencies; and c) The magnetic spectrum has more energy than the blackbody 
photon spectrum in vacuum, in particular, for frequencies, 
$\omega \leq 4\omega_{pe}$, where $\omega_{pe}$ is the electron plasma 
frequency.

\begin{center}
{\bf ACKNOWLEDGMENTS}
\end{center}
The authors would like to thank Swadesh Mahajan for useful suggestions, 
especially concerning the BGK collision term. The authors also would like to 
thank Arthur Elfimov for helpful discussions and the anonymous referee for 
useful comments and suggestions. M.O. would like to thank the Brazilian 
agency FAPESP for support and R.O. would like to thank the Brazilian agency 
CNPq for partial support.

\begin{figure}
\caption{The magnetic field spectrum 
$ln[S(\omega)/S_{0}]$ vs $\omega/\omega_{pe}$ (where 
$S(\omega)={\langle B^{2} \rangle}_{\omega}/8\pi$ 
and $S_{0}={\omega_{pe}}^{2}k_{B}T/c^{3}$ is the normalization)
for the {\it cold plasma} description with $x_{upper}=x_{cut}\sim 1$ for: 
(a) The electron-positron plasma at $T=7\times 10^{9}~K$ and 
$n_{e}=4.6\times 10^{30} cm^{-3}$
(the dashed curve is the {\it cold plasma} spectrum for low frequencies, 
the dash-dot-dash curve is the {\it cold plasma} spectrum for 
high frequencies (the link point between the two curves is indicated), and 
the solid curve is the blackbody spectrum in vacuum); 
(b) The same as case (a), where the {\it cold plasma} spectrum is 
plotted extended to high frequencies;  
(c) The same as case (a) but for an electron-proton plasma at 
$T=10^{9} K$ and $n_{e}=5.4\times 10^{26}$; and 
(d) The same as case (b) but for an electron-proton plasma at 
$T=10^{9} K$ and $n_{e}=5.4\times 10^{26}$.} 
\label{fig1}
\end{figure}
\begin{figure}
\caption{The magnetic field spectrum 
$ln[S(\omega)/S_{0}]$ vs $\omega/\omega_{pe}$ 
for the {\it cold plasma} description for various upper limits 
(where $S(\omega)={\langle B^{2} \rangle}_{\omega}/8\pi$ and 
$S_{0}=\omega_{pe}^{2}k_{B}T/c^{3}$ is the normalization). 
The plasma is an electron-positron plasma at 
$T=7\times10^{9}~K$ and $n_{e}=4.6\times 10^{30}~cm^{-3}$.  
The dash-dot-dash curve is the spectrum with $x_{upper}=x_{cut}$; the dotted 
curve is the spectrum with $x_{upper}=2~x_{cut}$; the dashed curve is the 
spectrum with $x_{upper}=5~x_{cut}$; and the dash-two dot-dash curve is
the spectrum with $x_{upper}=x_{max}$. The solid curve is the blackbody 
spectrum in vacuum.}
\label{fig2}
\end{figure}
\begin{figure}
\caption{The magnetic field spectrum 
$ln[S(\omega)/S_{0}]$ vs $\omega/\omega_{pe}$
(where $S(\omega)={\langle B^{2} \rangle}_{\omega}/8\pi$ 
and $S_{0}=\omega_{pe}^{2}k_{B}T/c^{3}$ is the normalization)  
for the {\it cold plasma} description with $x_{upper}=x_{max}$ for:  
(a) The electron-positron plasma  
at $T=7\times10^{9}~K$ and $n_{e}=4.6\times 10^{30}~cm^{-3}$ 
(the dashed curve is the {\it cold plasma} spectrum and the solid curve 
is the blackbody spectrum in vacuum); and 
(b) The same as case (a), extended to high frequencies showing the 
blackbody behaviour.}
\label{fig3}
\end{figure}
\begin{figure}
\caption{The magnetic field spectrum $ln[S(\omega)/S_{0}]$ 
vs $\omega/\omega_{pe}$ (where 
$S(\omega)={\langle B^{2} \rangle}_{\omega}/8\pi$ 
and $S_{0}=\omega_{pe}^{2}k_{B}T/c^{3}$ is the normalization)  
for {\it our model} for: (a) The electron-positron plasma at 
$T=7\times 10^{9}~K$ and $n_{e}=4.6\times 10^{30} cm^{-3}$
(the dotted curve is the spectrum of {\it our model} and the solid 
curve is the blackbody spectrum in vacuum); (b) The same as case (a), 
extended to high frequencies; (c) The same as case (a) but for an 
electron-proton plasma at $T=10^{9} K$ and $n_{e}=5.4\times 10^{26}$; 
and (d) The same as case (c) extended to high frequencies.} 
\label{fig4}
\end{figure}

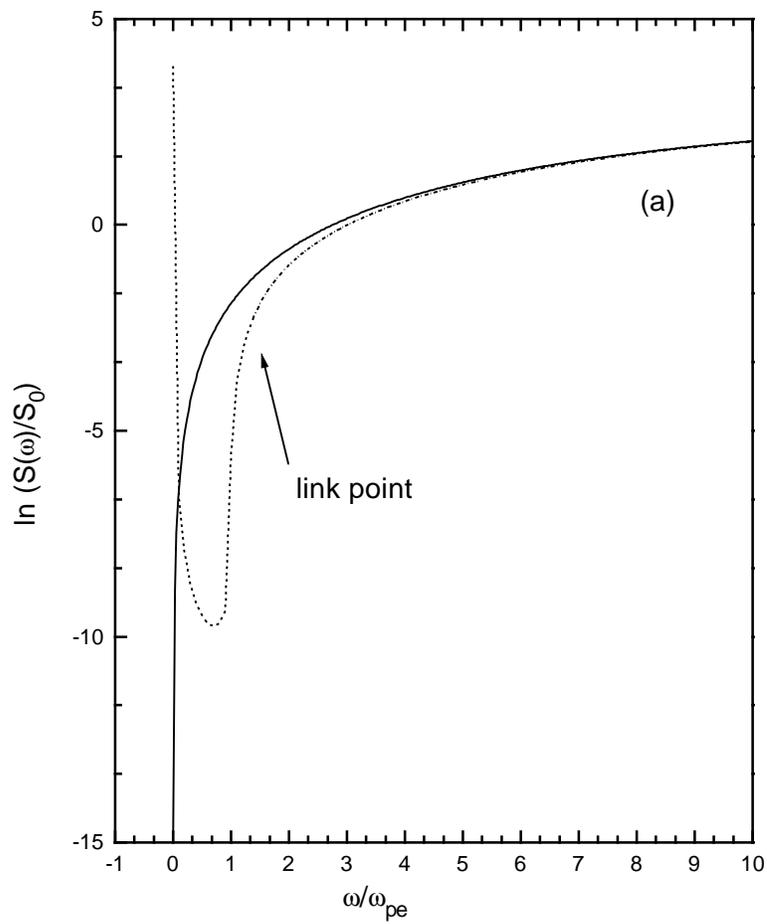
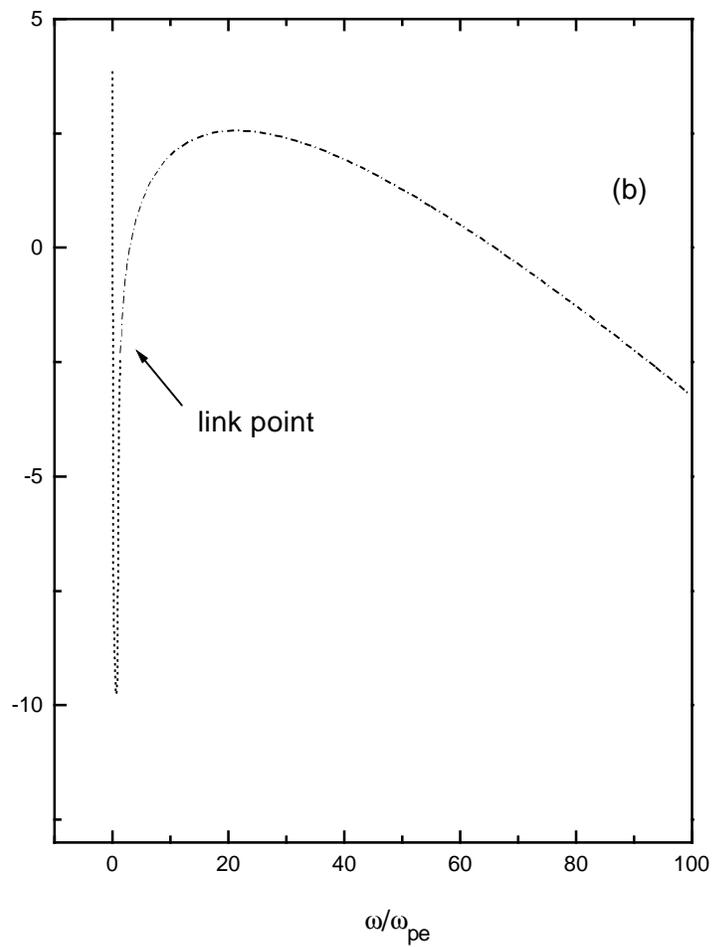
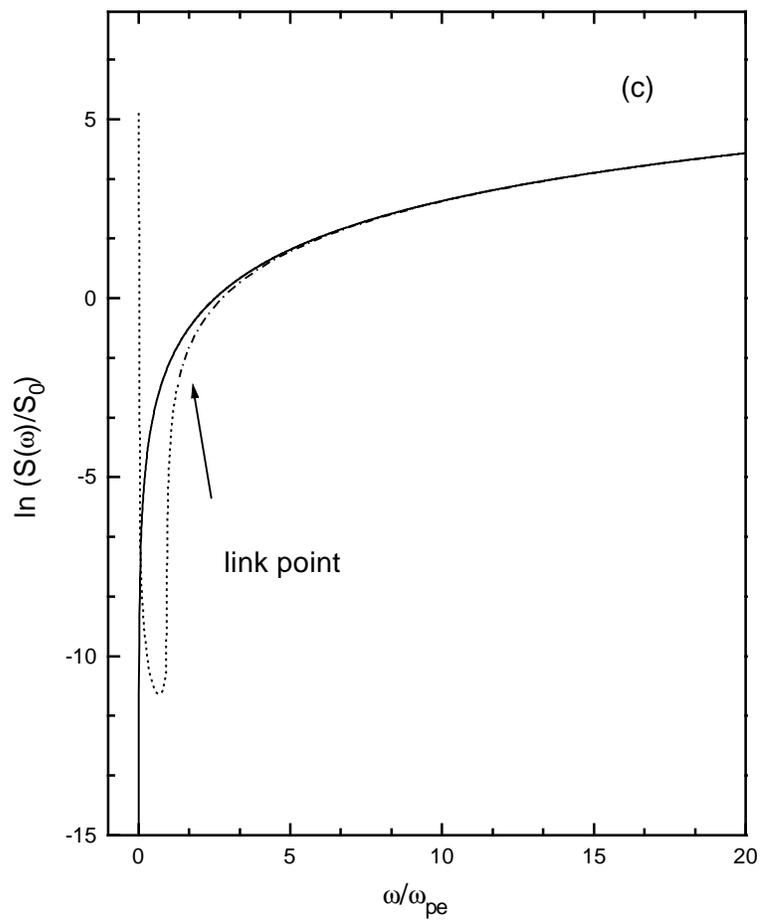
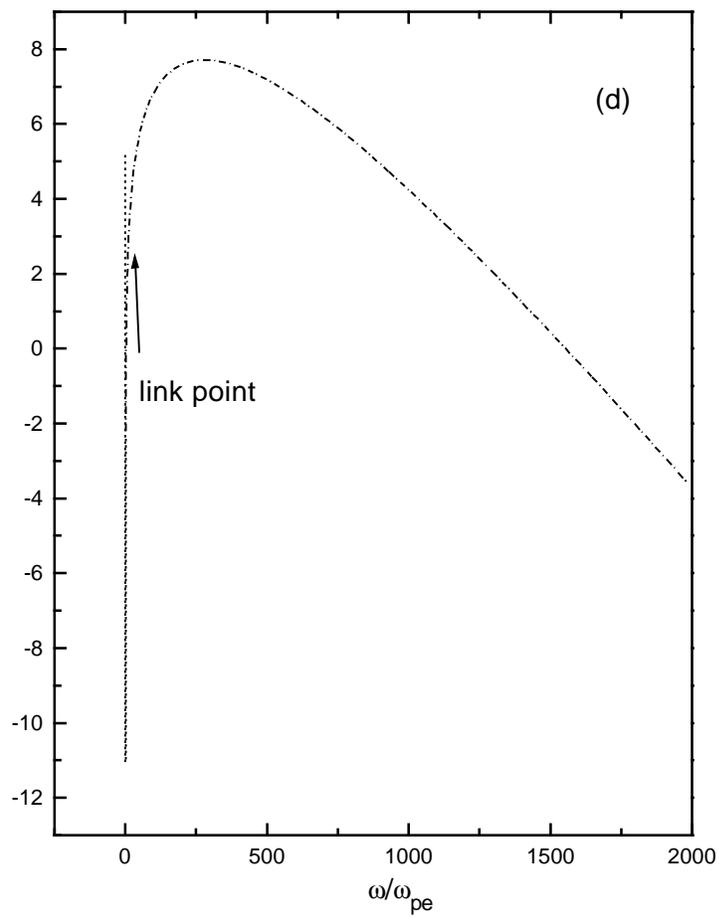
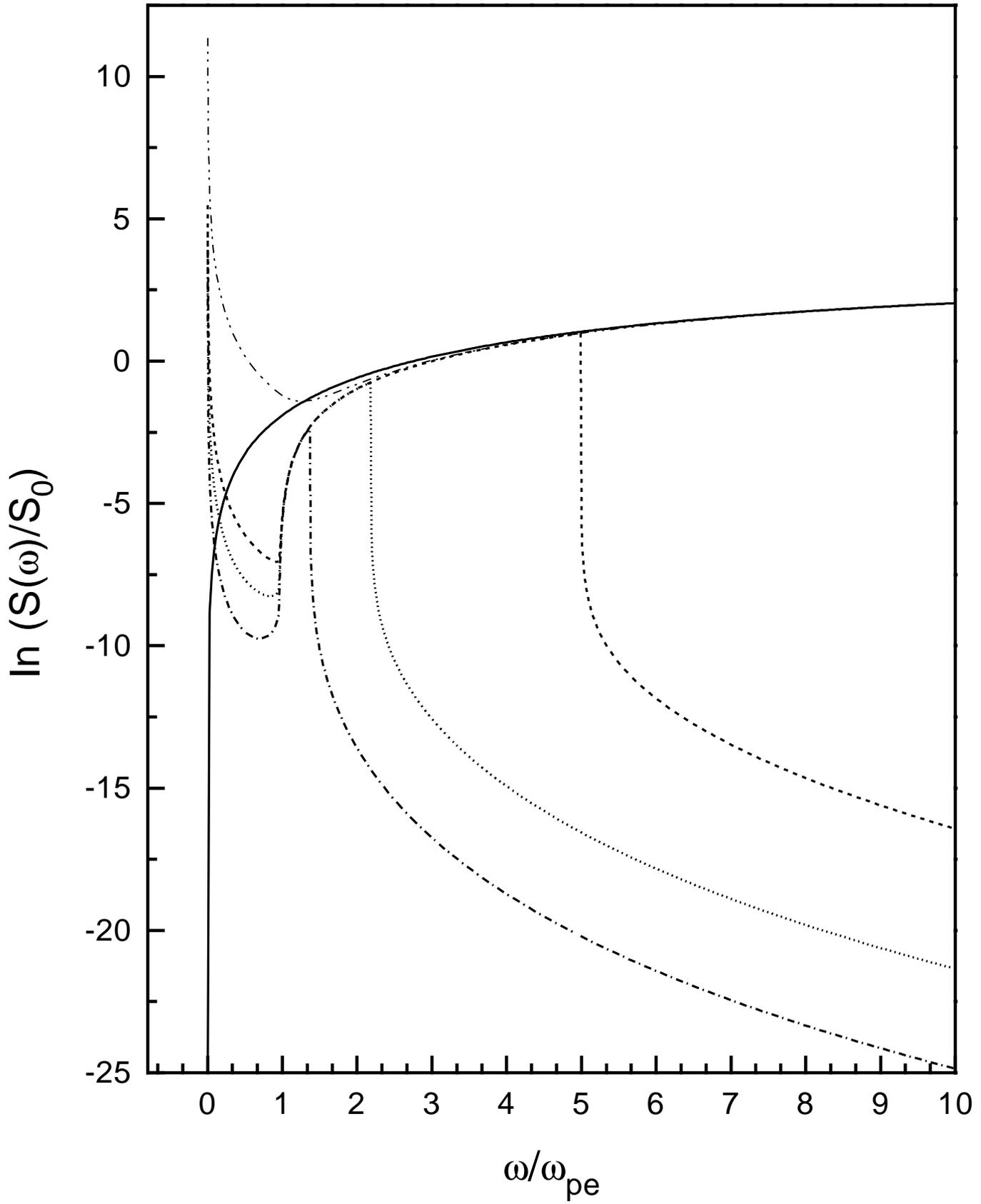
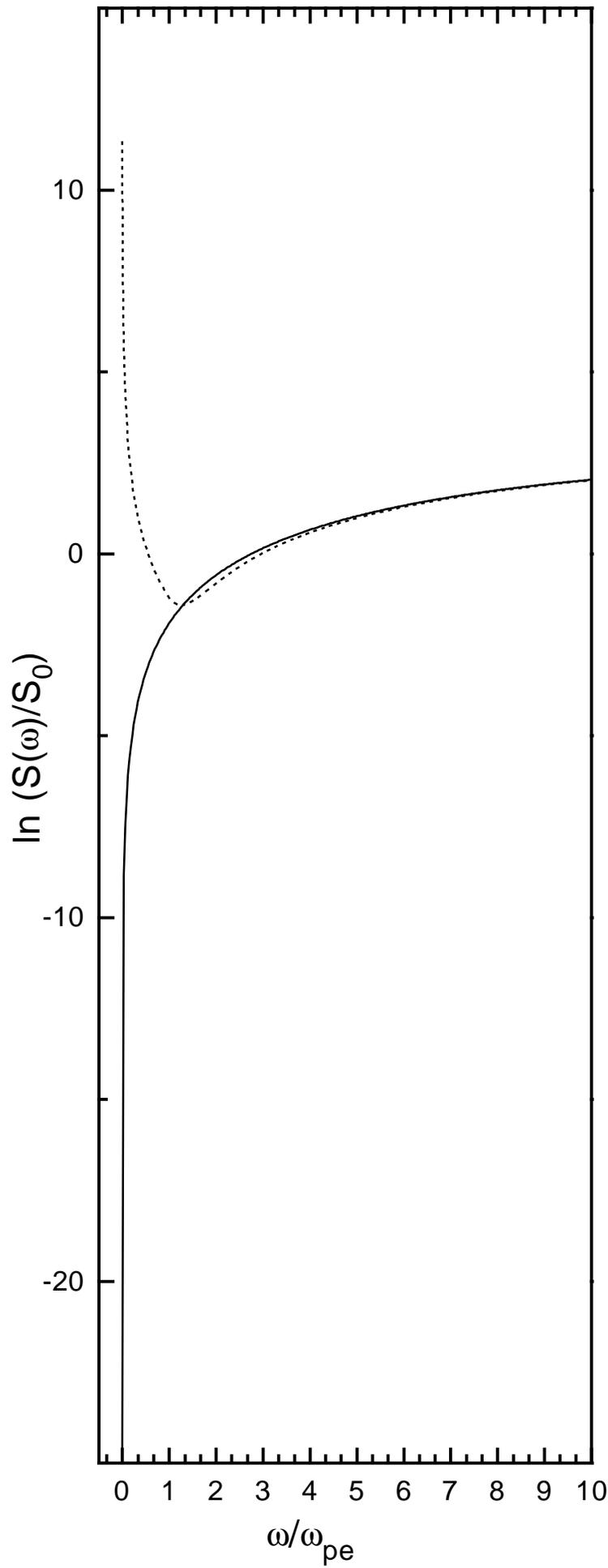
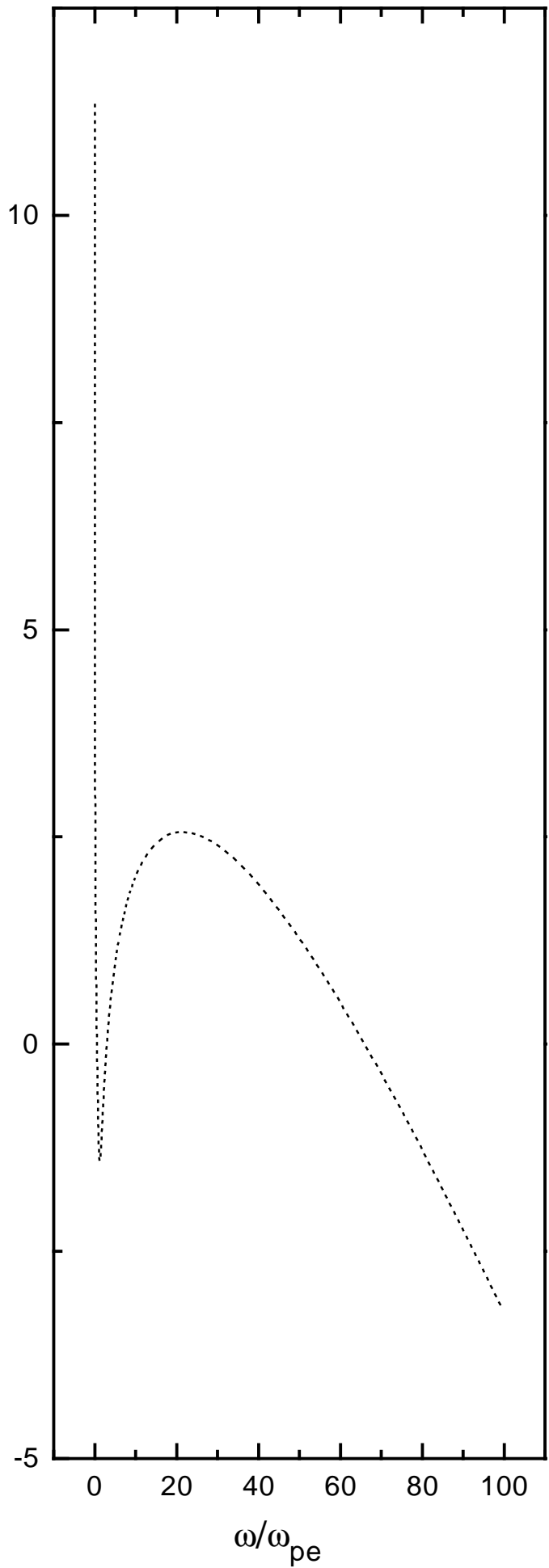
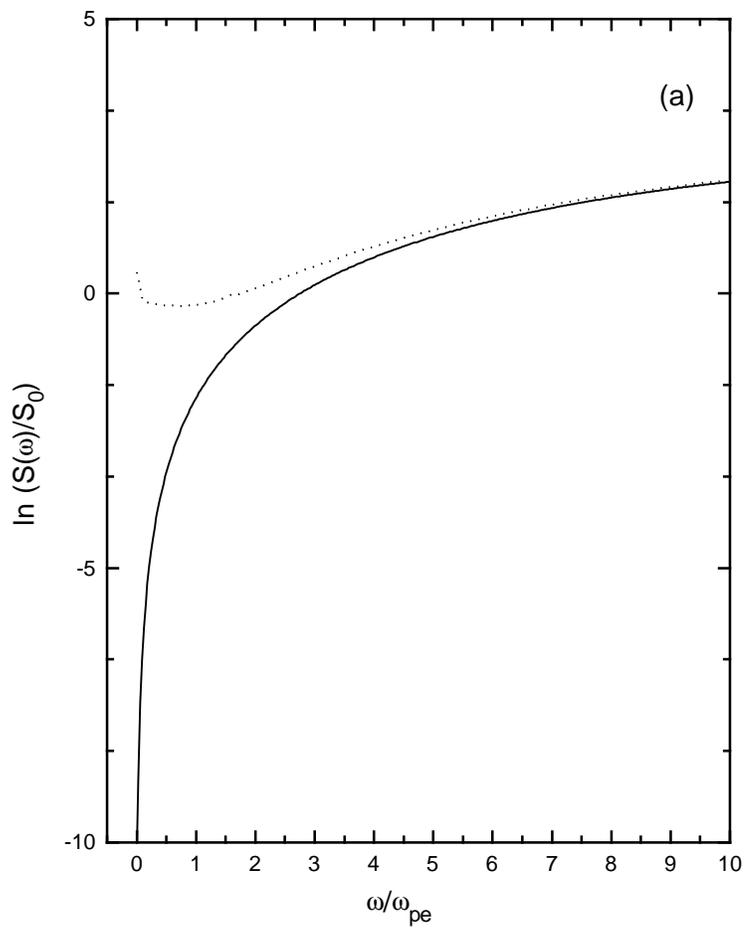
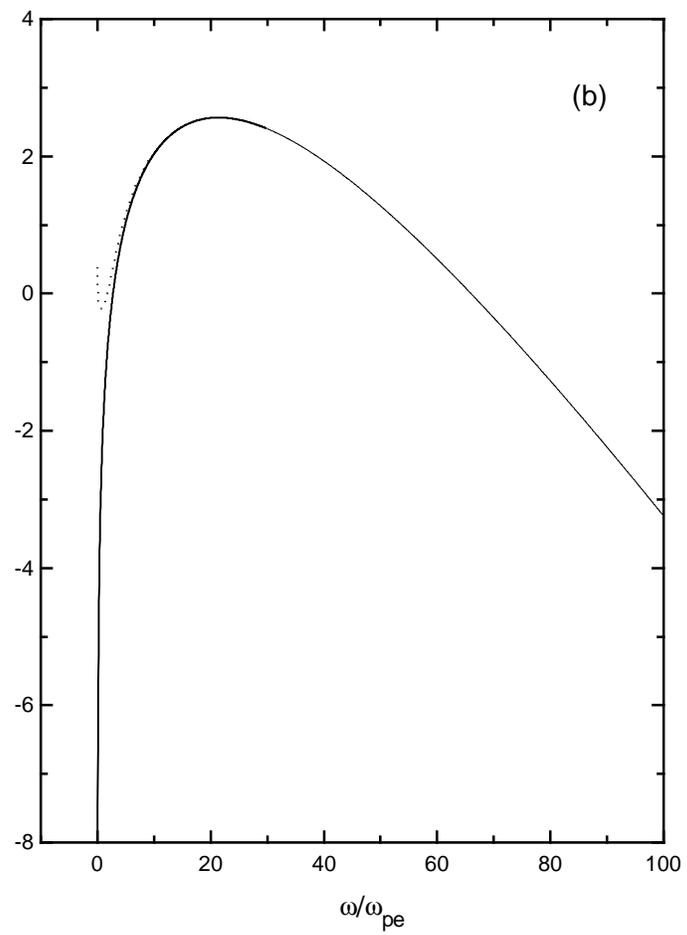
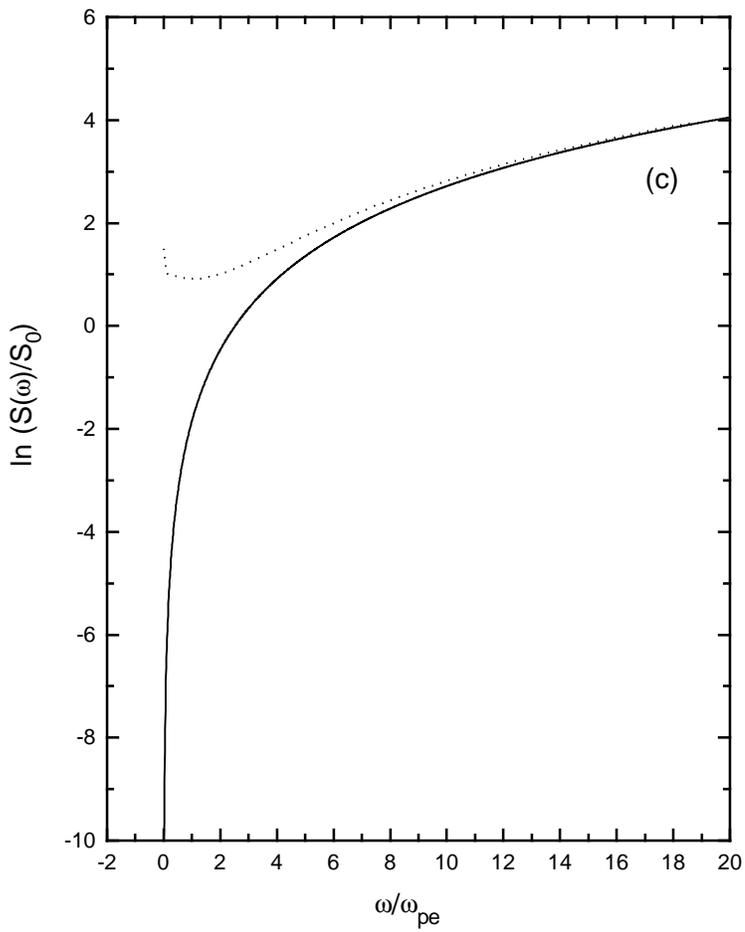
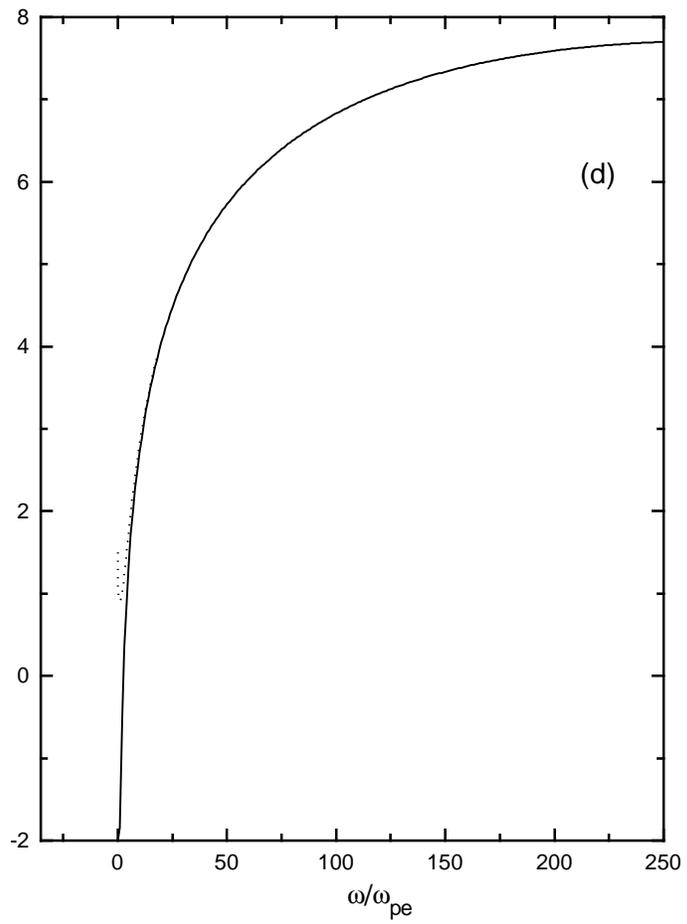
\end{document}